\documentclass[twocolumn,english,prx]{revtex4-2}
\usepackage{color}
\usepackage[utf8]{inputenc}
\usepackage{graphicx}
\usepackage[T1]{fontenc}
\usepackage{float}
\usepackage{natbib}
\usepackage{textcomp}
\usepackage{amsmath}
\usepackage{amssymb}
\usepackage{graphicx}
\usepackage{esint}
\usepackage{natbib}
\usepackage{xcolor}
\usepackage{multirow}
\usepackage{ulem}

\begin{document}
\title{Aharonov-Bohm oscillations in phosphorene quantum rings: mass anisotropy compensation  by confinement potential
}

\author{Tanmay Thakur and Bart\l{}omiej Szafran}

\affiliation{AGH University of Science and Technology, Faculty of Physics and
Applied Computer Science,\\
 al. Mickiewicza 30, 30-059 Kraków, Poland}

\begin{abstract}
We consider the Aharonov-Bohm (AB) effect on a confined electron ground state in a
 quantum ring defined electrostatically within the phosphorene monolayer. 
The strong anisotropy of effective masses in phosphorene quenches ground-state oscillations for a circular ring
 because of interrupted persistent current circulation around the ring. 
An elliptic deformation of the confinement potential can compensate
 for the anisotropy of the effective masses and produce ground-state
 parity transformations with the AB periodicity. 
Moreover, a specific ratio of the semiaxes is determined
 for which the spectrum becomes identical to that of a circular quantum ring and an isotropic effective mass.
 We identify a generalized angular momentum operator which commutes with the continuum Hamiltonian for the chosen ratio
 of the semi-axes that closes the avoided crossings of energy levels for states of the same parity and spin.
Ground-state oscillations for the two-electron ground state are also discussed.
\end{abstract}
\maketitle
\section{Introduction}

Phosphorene \cite{fosf14} or a monolayer form of black phosphorus \cite{bp,review0,rev} 
 is extensively studied for optics \cite{review}, field effect transistors \cite{bp,he19,bl19} and  quantum Hall effects \cite{hhe,ehe,qhe}.
Unlike the half-metallic graphene, phosphorene is a direct gap semiconductor that can host
the electrostatic lateral confinement of electrons. The electrostatic fields 
produce a clean confinement in gated two-dimensional systems
for investigation of the single-electron and interaction effects \cite{eqd} in carriers traps.
A particular form of the lateral confinement that attracts a lot of attention is the quantum ring \cite{fomin}. 
The annular confinement allows for  persistent current circulation in the presence of an external magnetic field, 
with the spectrum and magnetic response which is periodic with the
Aharonov-Bohm periodicity \cite{ab}. 
The periodicity of the spectrum of a phosphorene ring defined as a rectangular flake of the 
crystal with a central opening has been studied in Ref. \cite{pet} including the effect of the zigzag and armchair edges of the crystal. The purpose of this paper is to investigate
a clean quantum ring defined within phosphorene by an external potential that keeps the 
electrons off the edges of the crystal and is not affected by its details.

The anisotropy of the phosphorene crystal structure  \cite{rev} results in a strongly anisotropic 
electron effective mass \cite{bp,anisobp,18,tibikast1,tibikast2} that is much larger along the
zigzag chains of ions \cite{18} than in the perpendicular direction. The anisotropy prevents the persistent current flow in the electron ground state confined in a circular ring. However, current circulation
can be restored by deforming the confinement potential to an elliptic form.
Then, the spectrum acquires a braided pattern of even and odd parity energy levels, which cross
with the Aharonov-Bohm period.
Moreover, we propose a geometry for the elliptic confinement in which a modified angular momentum operator,
with one of the Cartesian coordinates rescaled, commutes with the Hamiltonian and the energy spectrum becomes similar to the one of an electron in a circular quantum ring with isotropic effective mass. For a confined electron pair interacting with the Coulomb potential, the operator no longer commutes with the Hamiltonian, but the Aharonov-Bohm oscillations of the ground-state energy appear for a tuned  confinement potential. 
 
\section{Theory}

\subsection{Tight-binding model}
\begin{figure}
 \includegraphics[width=1\columnwidth]{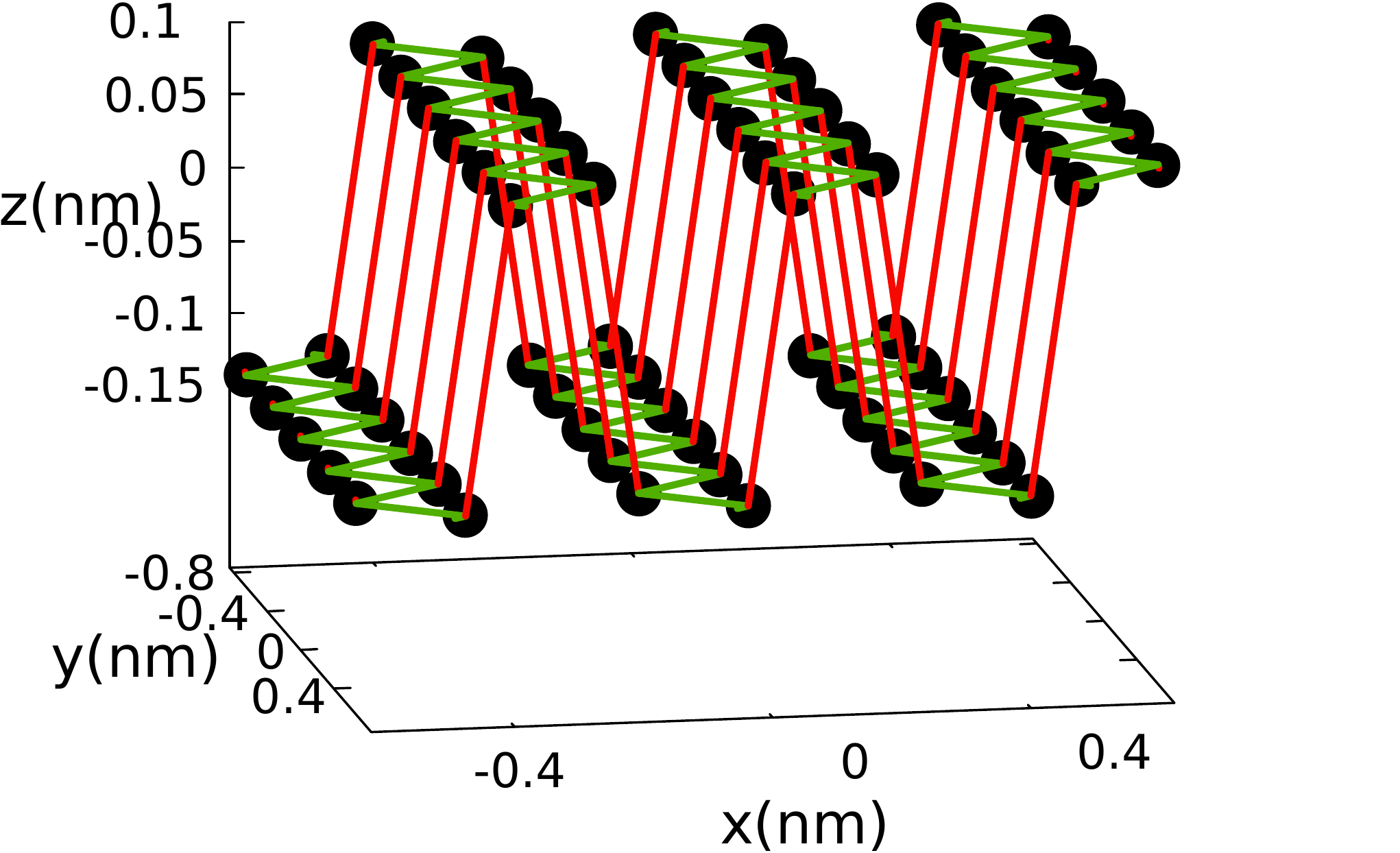}
\caption{Ions in phosphorene monolayer placed on two planes separated by a distance of 0.213 nm.
The lines link the neighbor ions with the
largest hopping energies (in-plane neighbors, green lines, hopping energy $-1.22$ eV) and (neighbors of separate planes,
red lines, hopping energy $3.665$  eV), see Table I.
}
 \label{crystal}
\end{figure}

We work with the phosphorene monolayer (see Fig. \ref{crystal}) using the Hamiltonian,
\begin{eqnarray}
H_{TB}=\sum_{kl } t_{kl} p_{kl} c_{k}^\dagger c_{l} 
 +\sum_{k} V_k c^\dagger_{k}c_{k}+{g\mu_B B}\sigma_z/2, \label{hb0}
\end{eqnarray}
where the first sum describes the hopping between the neighboring atoms.
The values for $t_{kl}$ (see Table \ref{tab1}) are taken from the five-parameter effective tight-binding Hamiltonian of Ref. \cite{tibikast1} . 
The positions of the ions in the phosphorene crystal \cite{review0} are plotted in Fig.1 
with the zigzag chains oriented along the $y$ direction.
In Eq. (\ref{hb0}),  $p_{kl}$ are the Peierls phase shifts that the electron acquires from the vector potential
along the line between $k$ and $l$ ions,
$p_{kl}=e^{i\frac{e}{\hbar}\int_{\vec{r_k}}^{\vec{r_l}}\vec {A}\cdot \vec {dl}}$.
We consider the magnetic field perpendicular to the monolayer $(0,0,B)$ with the vector potential taken in the symmetric gauge ${\bf A}=(-\frac{By}{2},\frac{Bx}{2},0)$.
In Eq. (\ref{hb0}), $V_k$ stands for the external potential on the ion $k$. The spin Zeeman effect is introduced by the last term of the Hamiltonian.\\

The $g$-factor for phosphorene of $g\approx 2.03$ was determined using $\textbf{k}\cdot\textbf{p}$ theory by Junior \textit{et al.} \cite{kp}. 
Zhou \textit{et al.} \cite{zhou} 
indicated the value of $g=2.14$ for monolayer black phosphorus. On the other hand, experiments have determined the value of $g=2 \pm 0.1 $ \cite{natyang} 
and $g\approx 1.8-2.7$ \cite{gatetune} 
and some have even reported $g=5.7\pm 0.7$ at low filling factors.
Nevertheless, the spin Zeeman term produces only a linear shift in the energy 
defining the spin splitting.
The magnetic field promotes the spin-down energy levels to the ground state
sooner or later on the magnetic field scale. We focus on avoided crossings
or crossings of the states of the polarized spin. 
The absence or presence of Aharonov-Bohm oscillations will not be affected by the value of the $g$-factor. Therefore, the spin Zeeman term is calculated with the value $g=2$.

\begin{table}
\begin{tabular}{l|r}

 $r_{kl}($nm$)$&  $t_{kl}($eV$)$  \\\hline   $0.222$&   -1.22 \\      $0.224$ &  3.665 \\ $0.334$ & -0.205\\$0.347$ &  -0.105 \\ $0.423$ & -0.055  \\ 
\end{tabular}
\caption{Hopping energies according to Ref. \cite{tibikast1} for a single black phosphorus layer.
The left column shows the distance between the ions and the right column the  hopping energy $t_{kl}$ applied in the tight-binding Hamiltonian (1).}
\label{tab1}
\end{table}

\subsection{Effective-mass Hamiltonian}
Part of the results of this work is obtained in the continuum approximation to the
tight-binding Hamiltonian. We use a single-band effective-mass operator,
\begin{eqnarray}
H_{em}&=&{\left(-i\hbar \frac{\partial }{\partial x}+e{A_x}\right)^2}/{2m_x}+{\left(-i\hbar \frac{\partial }{\partial y}+e{A_y}\right)^2}/{2m_y}\nonumber \\ &+&V(x,y)+{g\mu_B B}\sigma_z/2. \label{1eh}
\end{eqnarray}
with the effective mass parameters derived by fitting the tight-binding spectrum to the harmonic oscillator spectrum in Ref. \cite{szafran1} with
mass about five times heavier for the carrier motion along the zigzag chains of the crystal 
(see Fig. 1), $m_x=0.17037m_0$ and $m_y=0.85327m_0$.
The effective mass Hamiltonian is diagonalized using the finite-difference technique.

\subsection{Confinement potential}
	
We attempt to compensate for the anisotropy of the effective masses by the anisotropy of
the confinement potential. For that purpose, we use the following external potential
\begin{equation}
V(x,y)=\frac{1}{2}m_x\omega^2 (\rho(x,y)-R)^2,
\end{equation}
with $\rho(x,y)=\sqrt{x^2+y^2/\alpha}$, where $\alpha$ is a parameter that controls
the anisotropy of the potential. The potential vanishes
for points $(x,y)$ forming an ellipse 
\begin{equation}
\frac{x^2}{R^2}+\frac{y^2}{\alpha R^2}=1.
\end{equation}
We take the confinement energy $\hbar\omega=6$ meV. 
Changing $\alpha$ we keep the area within the ellipse fixed taking $R=\alpha ^{-1/4} R_c$, 
with $R_c=30$ nm, so that the number of magnetic flux quanta threading the ellipse is
the same for a given magnetic field $B$ independent of $\alpha$.
 For $R_c=30$ nm a flux quantum threads
the ring at 1.43 T, which is the period of the energy spectrum on the $B$ scale for a strictly 1D circular quantum ring. 
We use $\alpha\leq 1$ so that the half length of the major axis $a$ of the ellipse is oriented
along the $x$ axis $a=R=\alpha ^{-1/4} R_c$ and the half length
of the minor axis is $b=\sqrt{\alpha}R=\alpha^{1/4} R_c$.

In this work, we focus on the confined electron states of the conduction band.
In the continuum Hamiltonian, the bottom of the conduction band is set as the reference energy level.
The tight-binding spectrum produces  the conduction and valence
bands extrema spaced by the energy gap. For a finite flake, the spectrum contains
also in-gap states that are localized at the edge of the flake and the spectrum is not
symmetric with respect to the center of the energy gap \cite{pet}. 
Ref. \cite{pet} which used the same tight-binding parameterization \cite{tibikast1}
provides the lowest conduction-band state energy level of $\simeq 0.4$ eV
for a square flake with a side length of 8 nm in the absence of external potential.
In this paper, we are interested in states confined in the external potential that are independent of the details of the edge and thus correspond to an infinite crystal.
However, the calculations are carried out in an elliptical flake
for which the position of the ions satisfies the condition $x^2+y^2/\alpha<R_s^2$.
We take the flake large enough to contain all the discussed states within the confinement potential 
so that the results are independent of $R_s$. However, for $V=0$, the conduction band states occupy the entire flake, and the results  depend on $R_s$ \cite{reference}.
 The dependence on the lowest energy level in $R_s$ is well approximated by the dependence $E_0(R_s)=C/R_s^2+E_\infty$, where $C=673.31$(meV nm$^2$) and $E_\infty=340$ meV.
The $R_s^{-2}$ dependence is due to the finite-size effect, that is, the kinetic energy due to localization in a finite flake,
and $E_\infty=340$ meV is the estimated position of the bottom of the conduction band for an infinite crystal.
In the results presented in the following we shift down the tight-binding energies by $E_\infty$.

\section{Results and discussion}
\begin{figure}
\begin{tabular}{ll}
\includegraphics[height=0.25\textwidth]{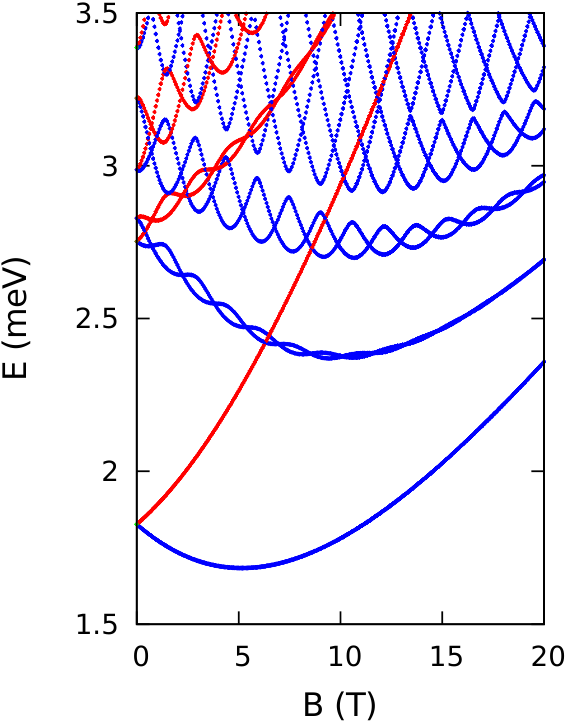} \put(-20,25){(a)}&\includegraphics[height=0.25\textwidth]{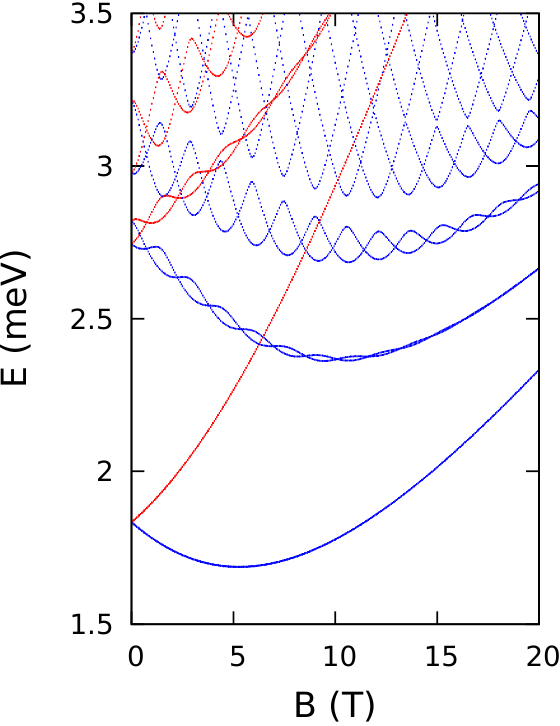}\put(-20,25){(b)} \\ \includegraphics[height=0.2\textwidth]{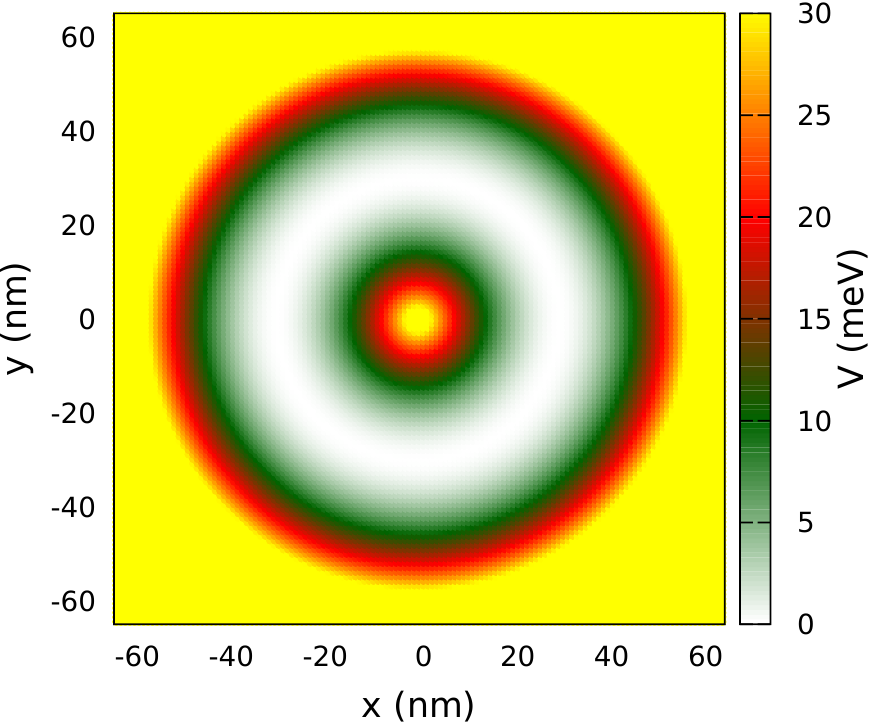}  \put(-35,25){(c)} &\includegraphics[height=0.2\textwidth]{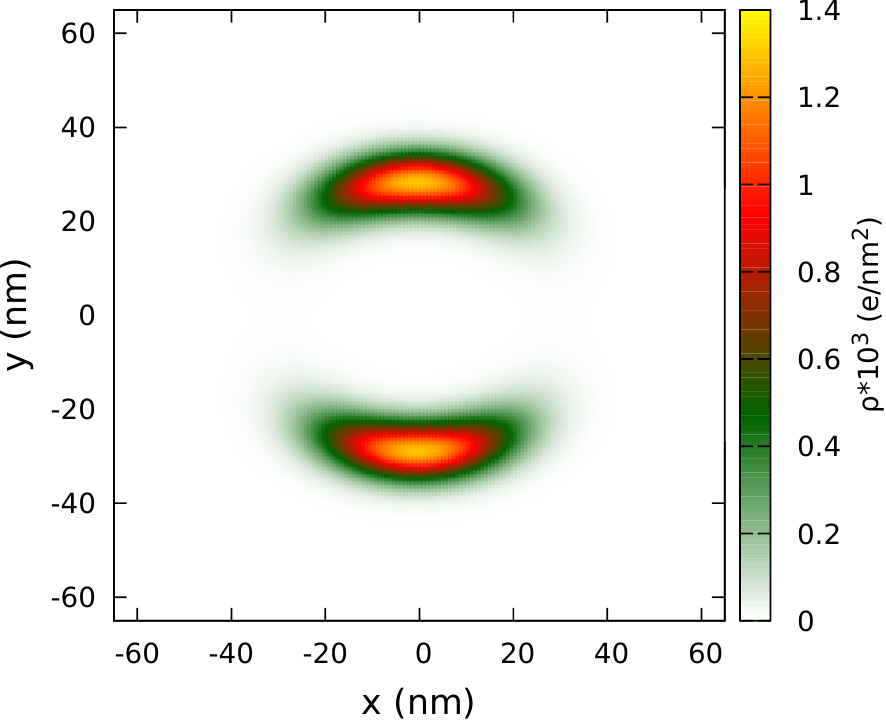}  \put(-40,25){(d)}\\
\end{tabular}
\caption{The tight binding (a) and continuum Hamiltonian (b) eigenvalues for $\alpha=1$. The blue and red lines show the spin-down and spin-up energy levels respectively. 
The tight-binding energy levels in (a) were shifted down by $E_\infty=340$ meV (see text in II.B).
Panel (c) show the confinement potential and panel (d) the ground-state charge density for $B=0$ calculated using the continuum model.}
\end{figure}

\begin{figure}
\begin{tabular}{ll}
\includegraphics[height=0.25\textwidth]{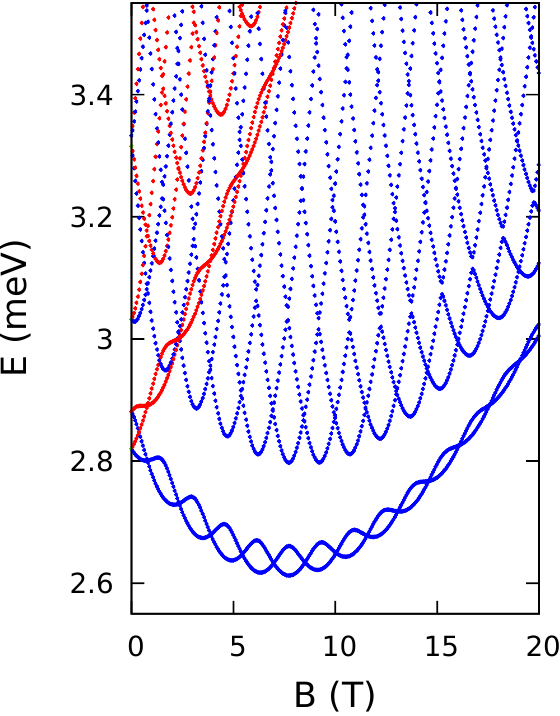} \put(-20,25){(a)} &\includegraphics[height=0.25\textwidth]{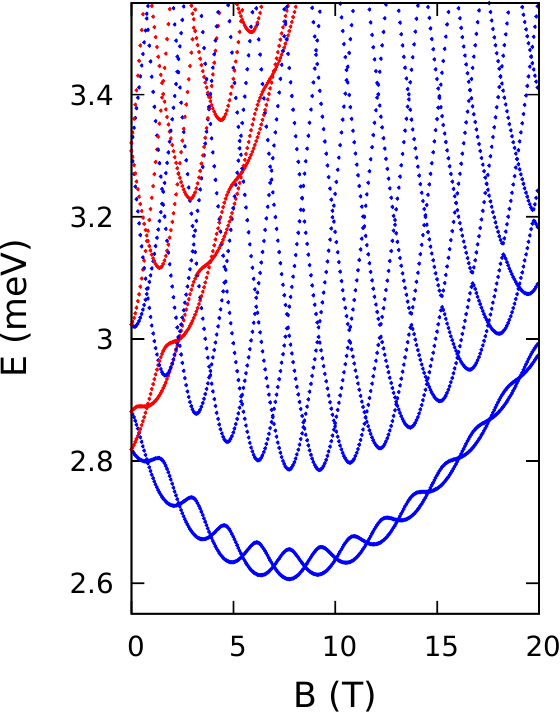} \put(-20,25){(b)}\\ \includegraphics[height=0.2\textwidth]{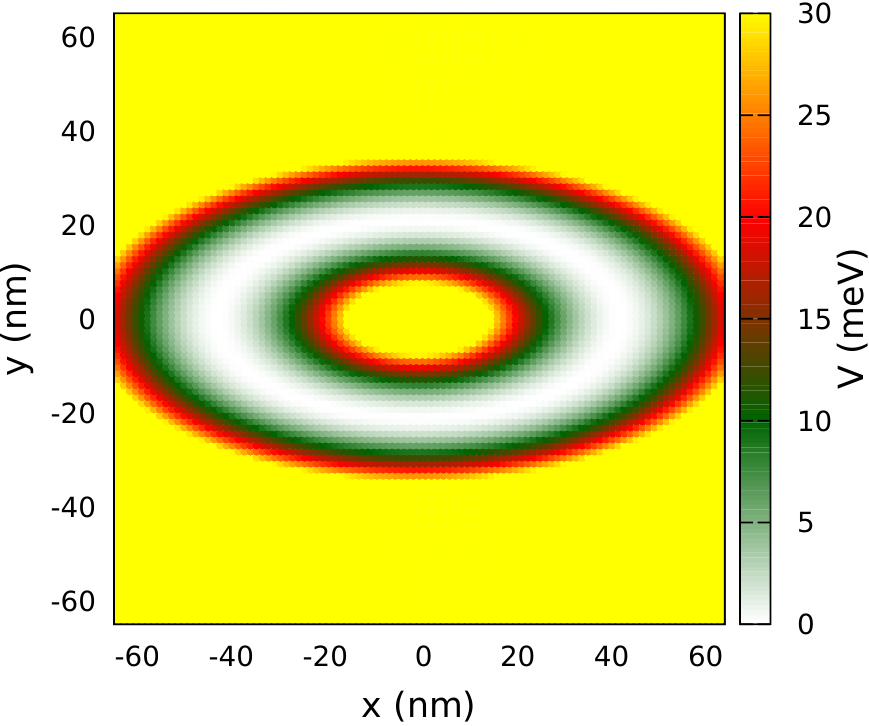}  \put(-35,25){(c)} &\includegraphics[height=0.2\textwidth]{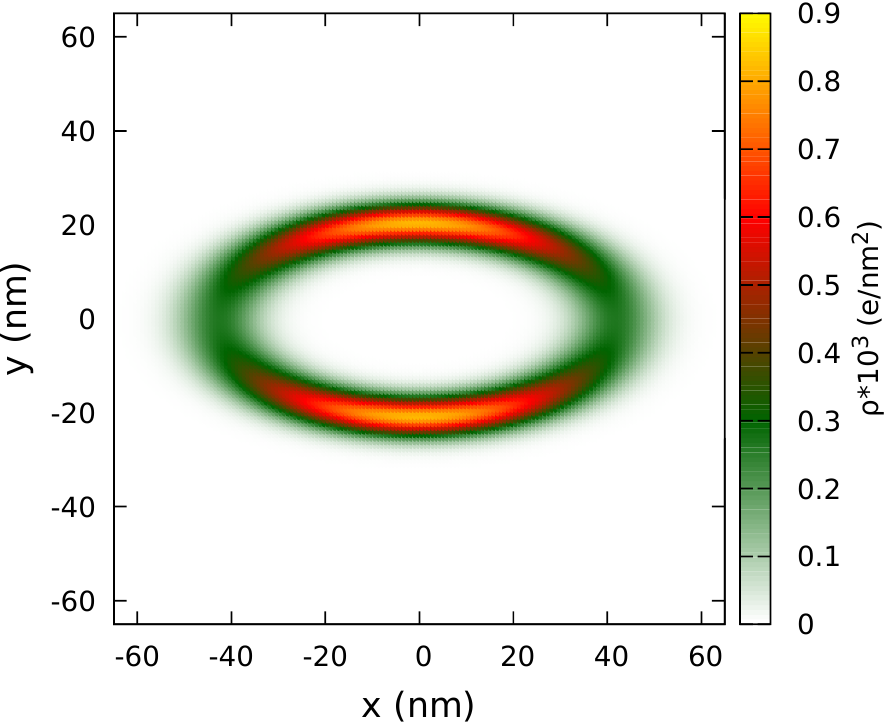}  \put(-40,25){(d)}\\
\end{tabular}
\caption{Same as Fig. 2 only for $\alpha=1.2{\frac{m_x}{m_y}}$. }
\end{figure}

\begin{figure}
\begin{tabular}{lll}
\includegraphics[height=0.25\textwidth]{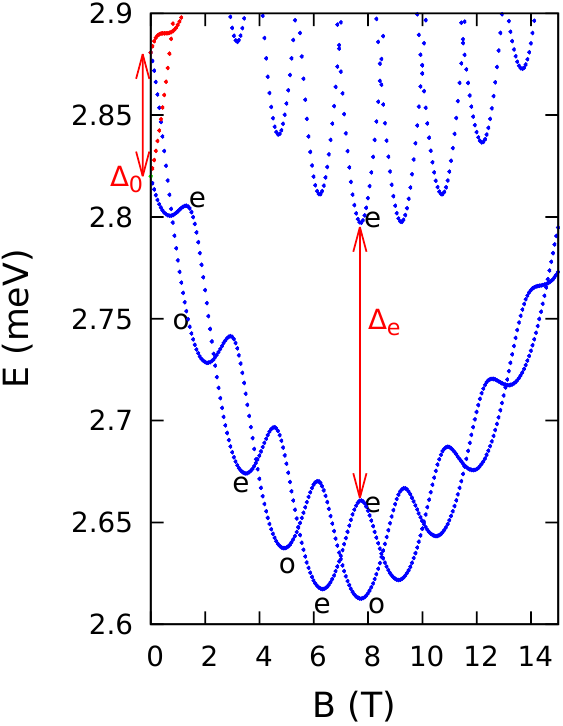} \put(-20,25){(a)}& \includegraphics[height=0.25\textwidth]{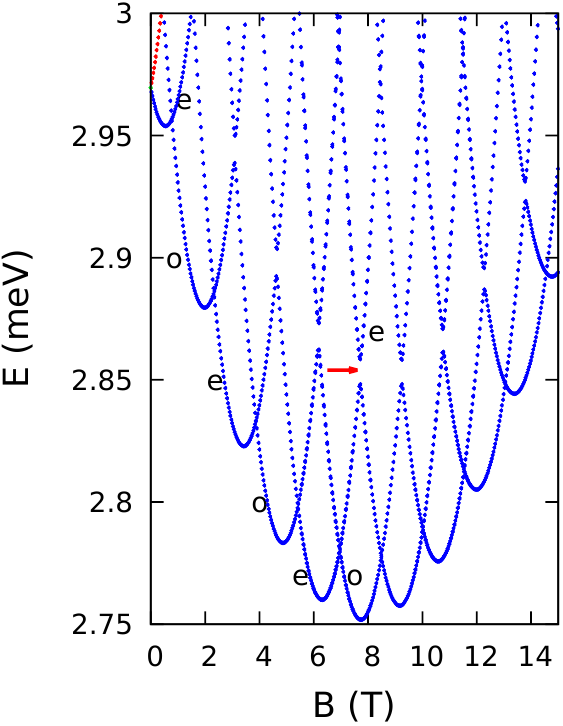}  \put(-20,25){(b)}
\end{tabular}
\caption{Magnified view of the low-energy part of the tight-binding spectrum for $\alpha=1.2{\frac{m_x}{m_y}}$ (a) [Fig. 4(a)]
 and $\alpha={\frac{m_x}{m_y}}$ (b) [Fig. 6(a)].
Letters 'e' and 'o' near the energy levels in (a) mark the even and odd parity energy levels
that are eigenstates of the parity operator with the eigenvalues $+1$ and $-1$, respectively.
$\Delta_0$ is the even-odd parity splitting at $B=0$ and $\Delta_e$ is the width of the avoided
crossing between the lowest even parity energy levels for $B\simeq 7.75$ T.
The red arrow in (b) shows the avoided crossing $\Delta_e$, here of the width of 13$\mu$eV.
(see Table II).}
\end{figure}

\begin{figure}
\begin{tabular}{ll}
\includegraphics[height=0.25\textwidth]{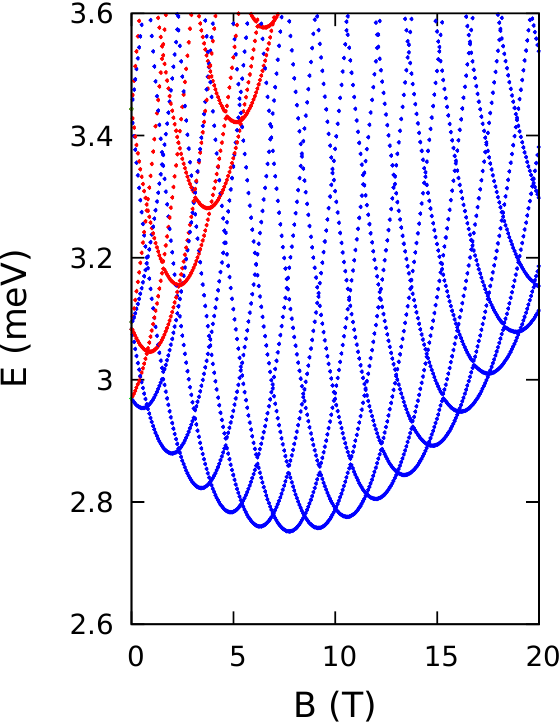} \put(-20,25){(a)}&\includegraphics[height=0.25\textwidth]{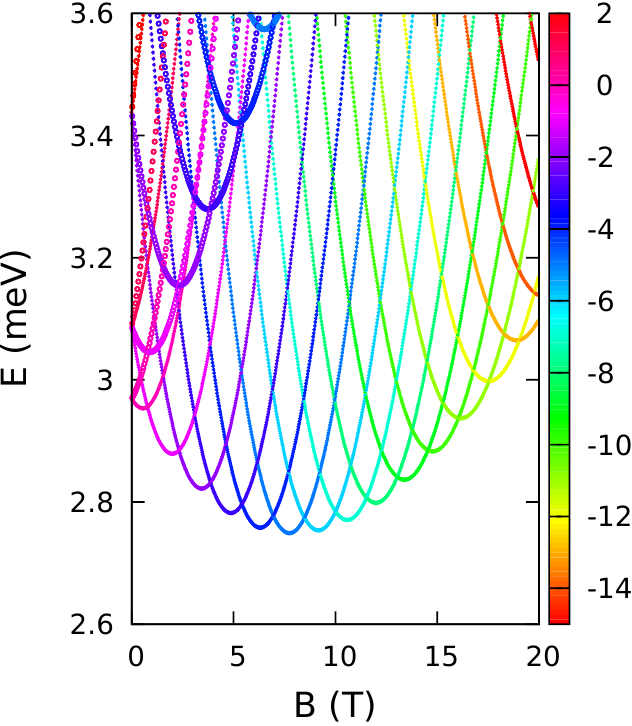} \put(-30,25){(b)}\\ \includegraphics[height=0.2\textwidth]{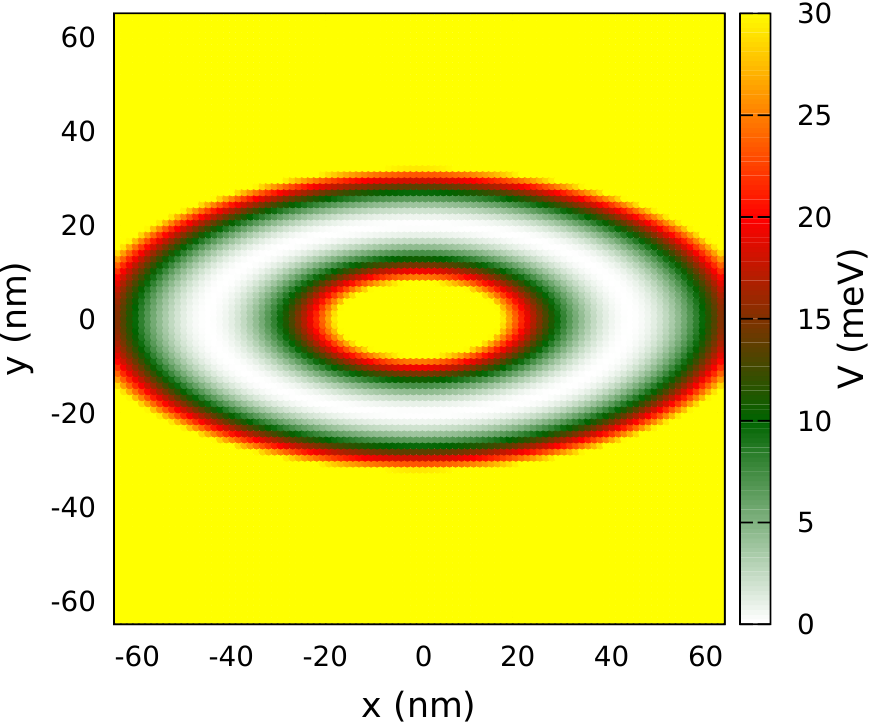} \put(-35,25){(c)}&\includegraphics[height=0.2\textwidth]{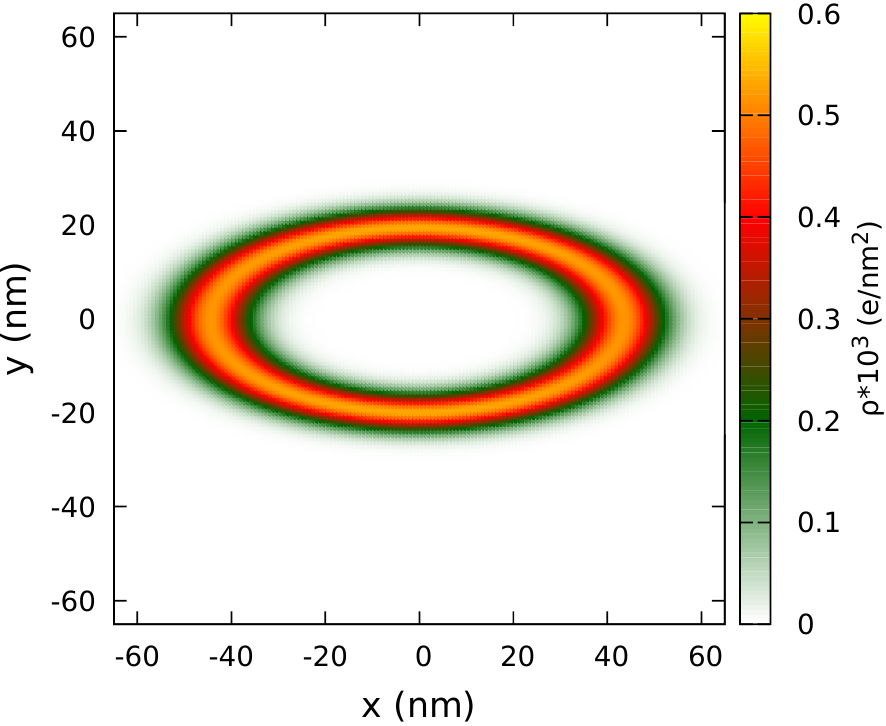} \put(-40,25){(d)}
\end{tabular}
\caption{Same as Fig. 2 only for $\alpha={\frac{m_x}{m_y}}$.
In panel (b) the colors indicate the eigenvalue of the angular momentum $l_z'$ operator,
and in panel (a) the colors describe the spin as in Fig. 2.}
\end{figure}

\subsection{Single-electron solutions}
The low-energy spectrum of a circular ring ($\alpha=1$) is plotted in Fig. 2(a) (Fig. 2(b)) for the tight-binding model (continuum model).
The TB results [Fig. 2(a)] here and below are shifted down by 340 meV.  
The results of both approaches are nearly identical.
The ground state of a circular ring is localized in two islands
near the $y$ axis on the opposite sides of the ring center  [Fig. 2(c)].
The variation of the wave function implies a contribution to the kinetic energy which is large
on the $x=0$ axis due to the low value of mass  $m_x$. In consequence the ground-state wave function
is far from the $x$ axis, and no persistent current circulation is possible in the ground-state.

The lowest spin-up and spin-down energy levels of Fig. 2(a,b) are two-fold degenerate with respect to the parity. Degeneracy results from the lack of tunnelling of the wave function across the $x$ axis (Fig. 2(d)).
In the excited state part of the spectrum, one can see pairs of energy levels that cross in a braid-like pattern.
The corresponding states have opposite parities, hence the crossings of the levels.
These oscillations are reminiscent of the angular momentum transitions for a circular
ring with  isotropic effective mass \cite{fomin,qrfb}.

With $\alpha<1$ the confinement area on the $y$ axis becomes thinner
and the one along the $x$ axis wider [see Fig.3(c) for $\alpha=1.2{\frac{m_x}{m_y}}$]. The confinement energy along the $y$-axis increases
and that along the $x$-axis decreases. The $x$ axis is now accessible for the ground-state electron  [Fig. 3(d)]. 
The ground-state degeneracy is lifted,
and the ground-state crossings of energy levels of opposite parity are observed [Fig. 3(a,b)].

In Fig. 4(a) we plotted a magnified view of the low-energy part of the spectrum. 
By $\Delta_0$ we denote the energy splitting of the lowest even and odd parity energy levels taken at $B=0$,
which defines the range of the ground-state energy oscillations as functions in the external magnetic field.
The braided two energy levels  cross with the Aharonov-Bohm period. The second quantity marked in Fig. 4(a)
by $\Delta_e$ is the width of the avoided crossing of even-parity energy levels taken at $B\simeq 7.75$ T.

Decreasing the anisotropy parameter to $\alpha={\frac{m_x}{m_y}}$ we find that the ground-state charge density [Fig. 5(d)] is constant along the confinement potential minimum Fig.  5(c)]. 
The continuum spectrum [Fig. 5(b)] contains crossings of energy levels in the entire spectrum. For the tight-binding
model -- see also Fig. 4(b) -- we find the parity-related crossings of the energy levels in the ground state as in Fig. 3, and only narrow avoided crossings are found
between the first and second excited energy levels of the same parity. The one marked by the red arrow in Fig. 4(b) corresponds
to $\Delta_e=13$ $\mu$eV. $\Delta_e$ attains its minimal value for $\alpha=\frac{m_x}{my}$.
Values of $\Delta_0$ and $\Delta_e$ calculated with the tight binding approach for varied $\alpha$ are summarized in Table II. 

\begin{table}
\begin{tabular}{l|l|l|l}
$\alpha$ & $\alpha/({m_x/m_y})$ & $\Delta_0$ (meV) & $\Delta_e$ (meV)\\ \hline 
1 & 2.238 &0.001 & 0.673 \\
0.894 & 2 & 0.012 & 0.414 \\
0.536 & 1.2 & 0.061 & 0.137 \\
0.469 & 1.05 & 0.096 & 0.046 \\
0.447 & 1 & 0.113& 0.013 \\ 
0.424 & 0.95 & 0.104& 0.028 \\ 
\end{tabular}
\caption{The spacing $\Delta_0$ between the lowest energy levels of even and odd parity for $B=0$ (see Fig. 4(a))
and the $\Delta_e$ width of the avoided crossing between the two lowest energy levels of even parity (Fig. 4(a)) as a function
of the eccentricity parameter $\alpha$ for $B\simeq 7.75$ T. 
The results are calculated with the tight binding approach.}
\end{table}

\subsection{Angular momentum in the rescaled space}
The crossings of the energy levels in the spectrum of Fig. 5(b)
suggest that an additional symmetry is present in addition to the parity.
With the substitution $y'={y}/\sqrt{\alpha}$ and $\alpha={\frac{m_x}{m_y}}$  the Hamiltonian (2) becomes
\begin{eqnarray}
H_{em}&=&-\frac{\hbar^2}{2m_x}\left(\frac{\partial^2}{\partial x^2}+\frac{\partial^2}{\partial y'^2}\right)+\frac{e^2B^2}{8m_y}(x^2+y'^2) \nonumber \\&+&\frac{eB}{2\sqrt{m_xm_y}}l_z'+V(\rho'-R)+{g\mu_B B}\sigma_z/2,
\end{eqnarray}
with $\rho'=\sqrt{x^2+y'^2}$ and
the $z$ component angular momentum operator in the deformed space $l_z'=i\hbar \left(y'\frac{\partial}{\partial x}-x\frac{\partial}{\partial y'}\right)$.
The confinement potential acquires circular symmetry upon rescaling of the $y$ coordinate and the Hamiltonian commutes with $l_z'$ operator.
The crossings of the eigenstates of the effective mass Hamiltonian are due to the symmetry which upon rescaling of the $y$ coordinate is no longer hidden.
The $l_z'$ eigenvalues are given by color in Fig. 5(b).  The ground state undergoes $l_z'$ angular momentum transitions
similar to the ones found for circular quantum rings with isotropic electron effective mass \cite{qrfb}.
Note that the applicability of $l'_z$ operator is not limited to quantum rings, but it can also be used to any potential profile which is radially symmetric for the rescaled $y$ coordinate.

\subsection{1D limit}

For a narrow radial confinement  (large $\omega$) the 
low-energy part of the spectrum occupy the same state of radial quantization and there is essentially one degree of freedom of motion along the ring.
The Hamiltonian (5) put in circular coordinates  reads
\begin{eqnarray}
H_{em}&=&-\frac{\hbar^2}{2m_x}\left(\frac{1}{\rho'}\frac{\partial }{\partial \rho'}+\frac{\partial^2 }{\partial \rho'^2}-\frac{ l_z'^2}{\rho'^2}\right)+\frac{B}{2\sqrt{m_xm_y}}l_z'+\nonumber \\ && +\frac{e^2B^2}{8m_y}\rho'^2+V(\rho'-R),
\end{eqnarray}
where we neglected the spin Zeeman term.
For strong confinement (large $\omega$) the radial profile of the wave function no longer depends on $l_z'$ or $B$.
Then, the terms with the derivatives with respect to $\rho'$ and the external potential produce the same energy contribution for all the states involved.
With this contribution set as the reference energy level, we obtain the energy spectrum of the form,
\begin{equation}
E(l_z',B)=\frac{\hbar^2l_z^2}{2m_xR^2}+\frac{e^2B^2}{8m_y}R^2+\frac{eB}{2\mu}l_z',
\end{equation}
where $\mu=\sqrt{m_xm_y}$.
With $R=\left(\frac{m_y}{m_x}\right)^{1/4} R_c$ one obtains an expression that is symmetric in the effective masses,
\begin{eqnarray}
E(l_z',B)&=&\frac{\hbar^2l_z'^2}{2\mu R_c^2}+\frac{e^2B^2}{8\mu}R_c^2+\frac{eB}{2\mu}l_z' \nonumber\\ &=&
\frac{\hbar^2}{2\mu R_c^2}\left(\frac{\Phi}{\Phi_0}+l_z'\right)^2, 
\end{eqnarray}
where $\Phi_0=\frac{h}{e}=\frac{2\pi\hbar}{e}$ is the flux quantum and $\Phi=B\pi R_c^2$.
The final result with the geometric average of the effective masses is identical to that of the circular ring with an isotropic effective mass \cite{qrfb}.

Figure 6 shows the 2D continuum Hamiltonian (6) spectra with the spin Zeeman effect excluded
for $\hbar\omega=6$ meV in Fig. 6(a), $\hbar\omega=120$ meV in Fig. 6(b) and the results of the 1D formula in Fig. 6(c).
In Fig. 6(a) we see a diamagnetic shift of the spectrum to higher energy. The 
period of the Aharonov-Bohm oscillations is slightly larger than in the 1D results 
due to compression of the wave function by the external magnetic field that decreases
average $\rho'$ below $R_c$. For the radial
wave function confined stronger around $R_c$
in Fig. 6(b), we see the results approach the results for the analytical formula (8).

\begin{figure}
\begin{tabular}{l}
\includegraphics[height=0.15\textwidth]{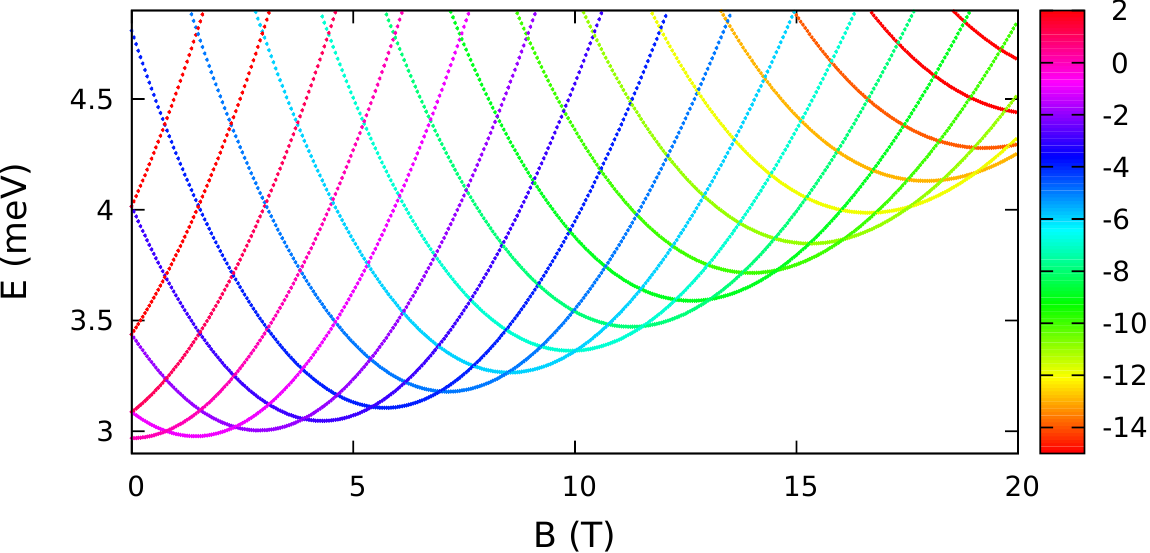} \put(-30,25){(a)}\\\includegraphics[height=0.15\textwidth]{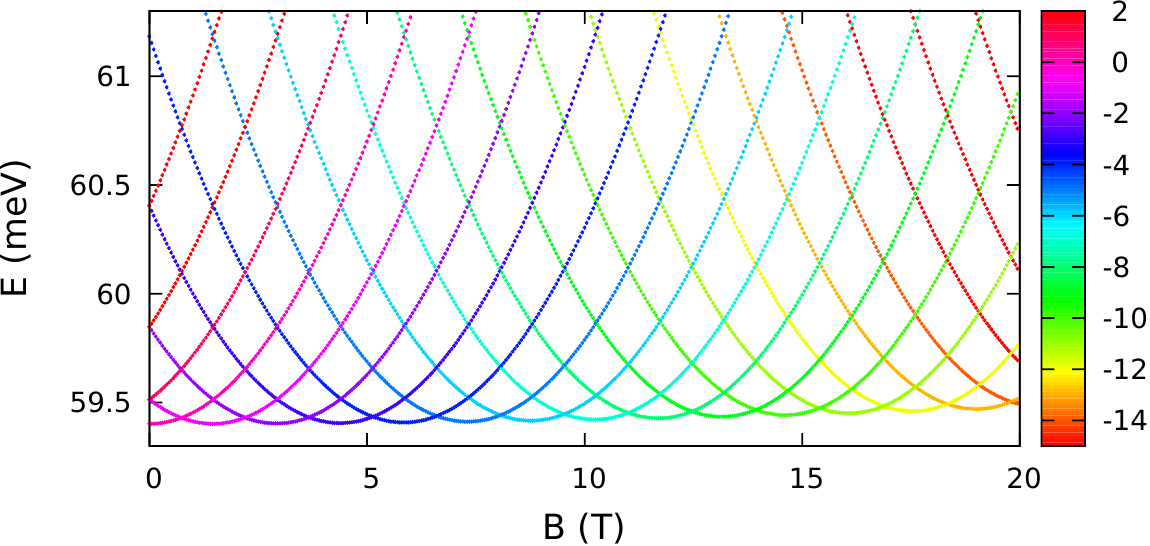} \put(-30,25){(b}\\ \includegraphics[height=0.15\textwidth]{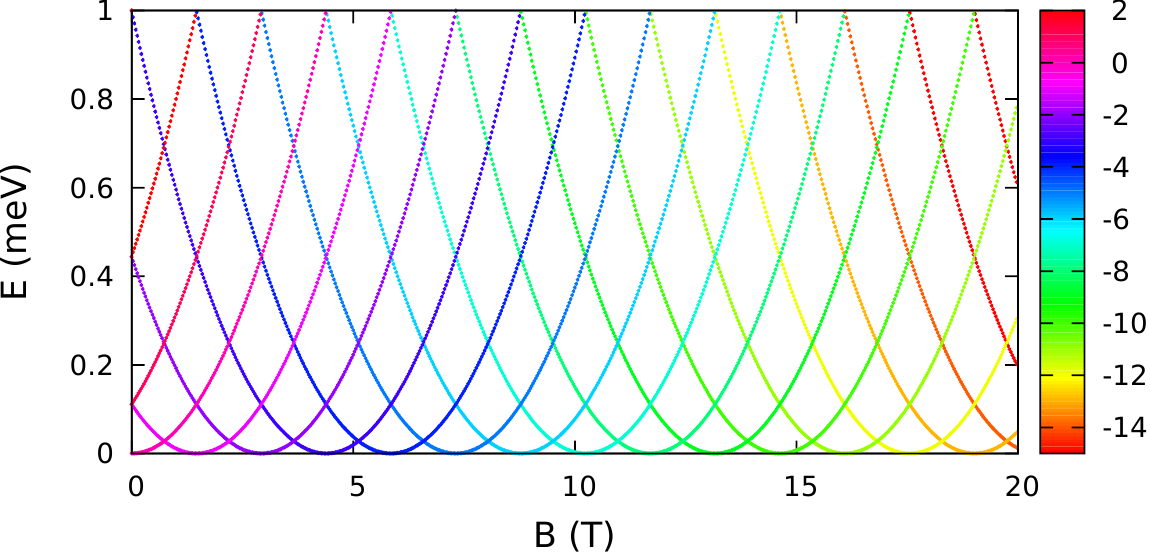} \put(-30,25){(c)}\end{tabular}
\caption{
The energy spectrum of Hamiltonian (5) with $g=0$ for $\hbar\omega=6$ meV (a), $\hbar\omega=120$ meV (b), and the 1D formula (7). The color of the lines gives the $l_z'$ operator eigenvalue}
\end{figure}

\subsection{Two-electron spectrum}
The electron-electron interaction potential does not commute with the $l_z'$ operator,
so the presence of the ground-state oscillations is not given a priori.
We calculated the two-electron energy spectrum in the continuum approach
using the Hamiltonian
\begin{equation}
H_{2e}=H_{em}({\bf r}_1)+H_{em}({\bf r}_2)+\frac{e^2}{4\pi\epsilon_0\epsilon} \frac{1}{r_{12}},
\end{equation}
where $H_{em}$ is the single-electron Hamiltonian (2), and $\epsilon=12$ 
is taken. The two-electron Hamiltonian is diagonalized in the basis of up of the two-electron Slater determinants
constructed from the 30 lowest-energy single-electron eigenfunctions of Hamiltonian (2).

For both $\alpha=1$ (Fig. 7(a)) and $\alpha={\frac{m_x}{m_y}}$ (Fig. 7(c)) at $B=0$
the ground state is four-fold degenerate, with singlet and triplet energy levels
of the same energy. The exchange interaction is zero since the electrons form
 single-electron islands that are completely separated. 
For $\alpha=1$ the localization of the charge density formed a single-electron island
already without interaction (Fig. 2(d)). For $\alpha={\frac{m_x}{m_y}}$ the electron-electron
interaction separates the electron density to the opposite ends of the longer semiaxis of the ellipse (Fig. 7(d)). A more or less uniform electron distribution is obtained for 
$\alpha=1.7{\frac{m_x}{m_y}}$ (Fig. 7(f)). 
In this case, the ground state for $B=0$ is a singlet which is not degenerate with the triplet.
The  field of about 0.5 T promotes the triplet to the ground state. 
Periodic avoided crossings are observed in the ground state, which are similar to the ones found
for a system with isotropic effective mass but in an anisotropic quantum ring \cite{fomin2}.
 For the circular ring, the triplet-energy levels
correspond to odd values of the total angular momentum ($L$) \cite{jlzhu} that correspond to the negative
parity $(-1)^L$. In our two-electron system, only the parity is a good quantum number. 
In Fig. 7(e) we see a series of avoided crossings between the ground state and
the first excited state. The avoided crossings obtained for the spin-polarized two-electron levels
are observed due to the same -- odd -- parity of these levels, which is in contrast to the 
single-electron ground state where the crossings of  energy levels corresponding to opposite parity
were observed in the ground state.

\begin{figure}
\begin{tabular}{ll}
\includegraphics[height=0.22\textwidth]{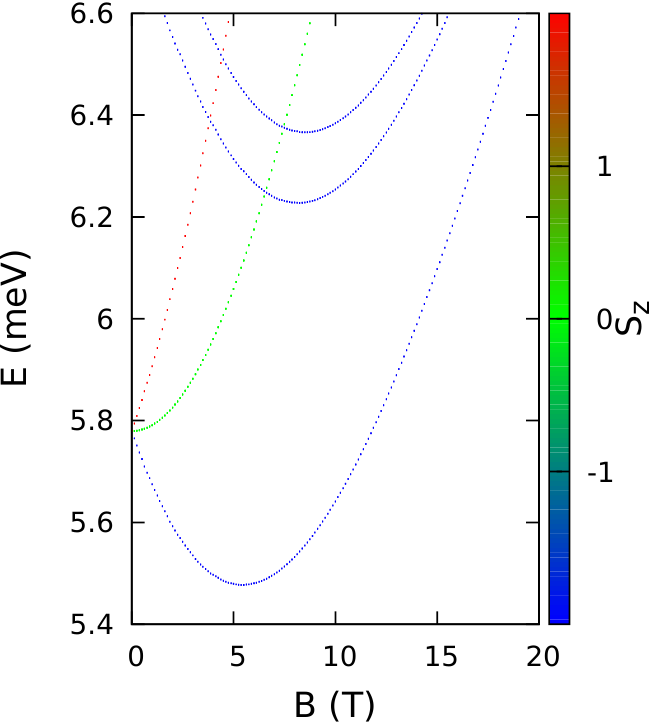} \put(-35,25){(a)} & \includegraphics[height=0.22\textwidth]{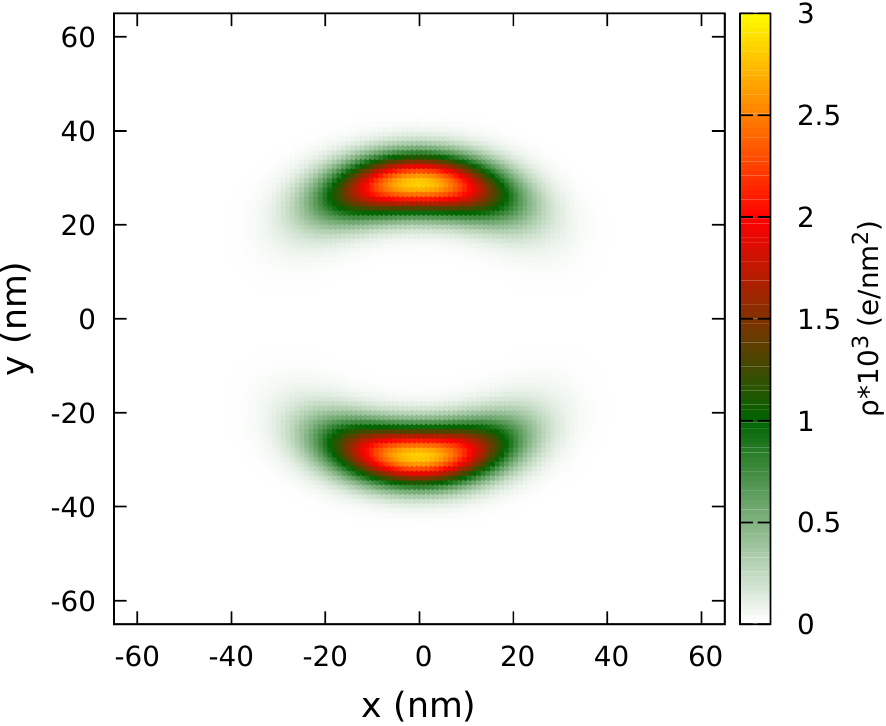} \put(-50,25){(b)}\\ 
\includegraphics[height=0.22\textwidth]{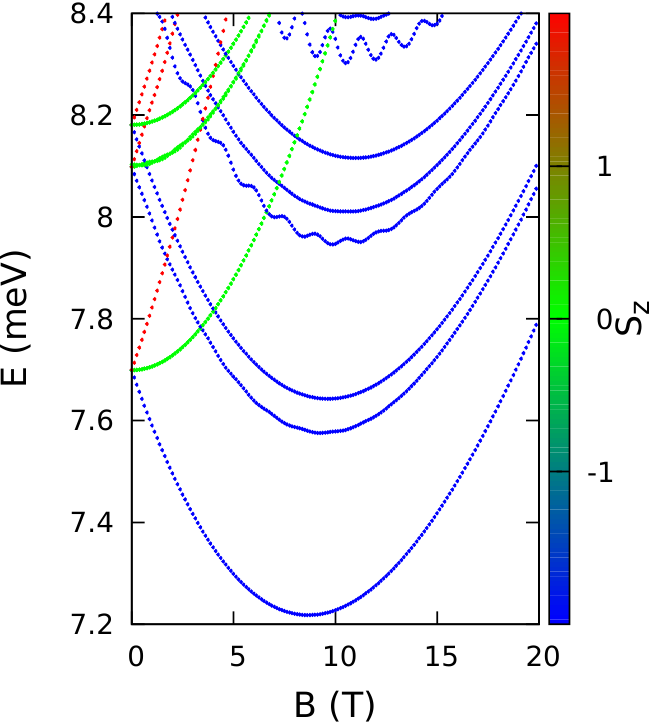} \put(-35,25){(c)}&
\includegraphics[height=0.22\textwidth]{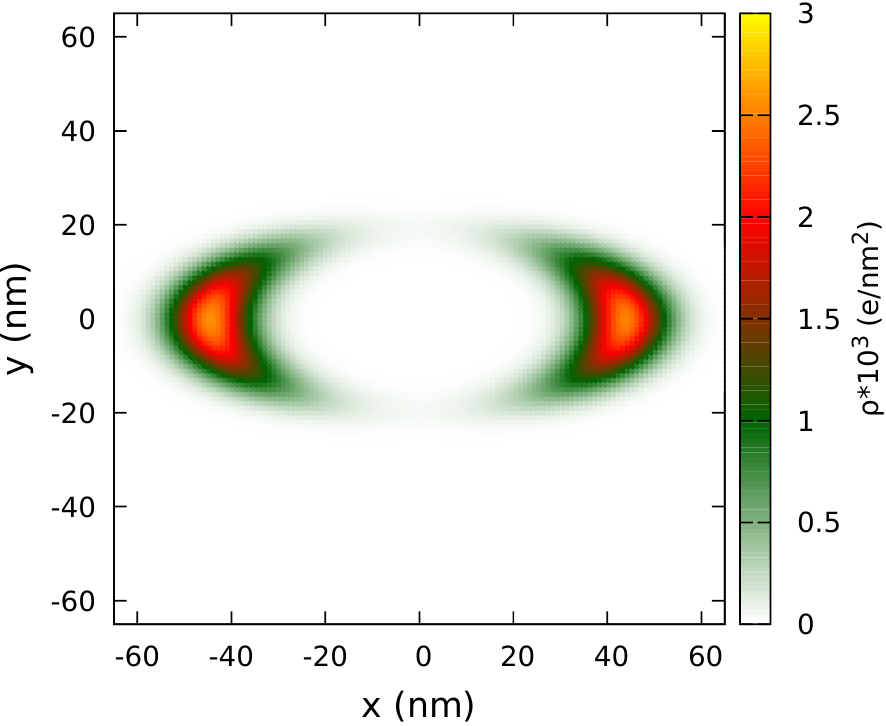} \put(-50,25){(d)}\\
\includegraphics[height=0.22\textwidth]{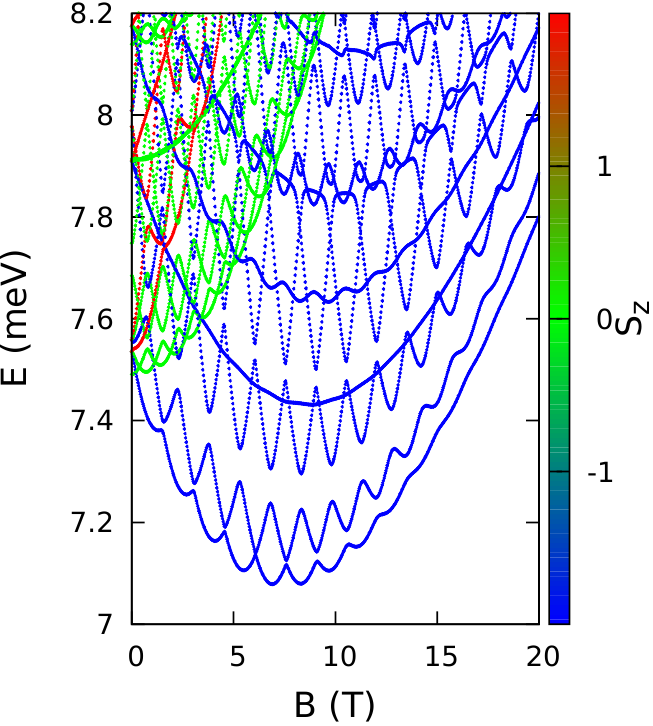} \put(-35,25){(e)} & \includegraphics[height=0.22\textwidth]{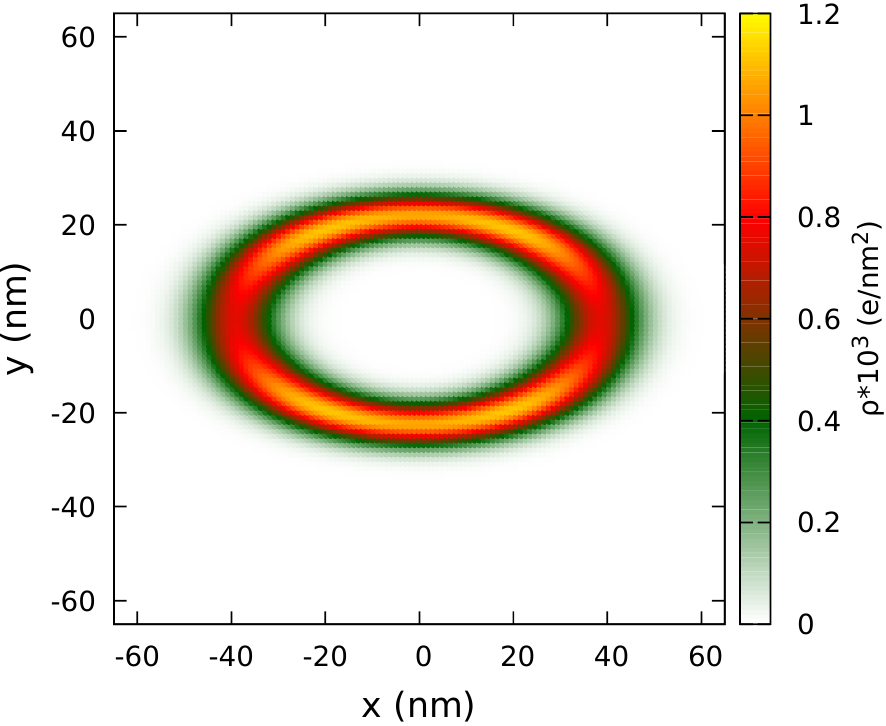} \put(-50,25){(f)}
\end{tabular}
\caption{The two-electron energy spectrum (a,c,e) and
the ground-state charge density (b,d,f) calculated using continuum model for $B=0$ and $\alpha=1$ (a,b), $\alpha={\frac{m_x}{m_y}}$ (c,d) and for $\alpha=1.7{\frac{m_x}{m_y}}$(e,f).
In (a,c,e) the color of the lines corresponds to the $z$ component of the total spin.}
\end{figure}

\subsection{Discussion}
The spectra of confined quantum rings are experimentally studied for annular-shaped  traps defined in gated two-dimensional electron gas \cite{fuhrer}. The current transport for electron traps weakly coupled to the electron reservoirs 
is governed by the Coulomb blockade \cite{leo,eqd}, with the single-electron current passing through the system
only when the chemical potential of the confined $N$-electron system falls within the transport window 
defined by the Fermi levels of the source and drain. The technique can also be used for detection of the
excited part of the spectra when the corresponding energy level enters the transport window \cite{leo,fuhrer}. 
The transport spectroscopy allows for reconstruction of the energy spectra with
a precision of the order of a few $\mu$eV \cite{fuhrer} with the Aharonov-Bohm periodicity as the signature
of the angular confinement.
The gated quantum rings can exhibit  an elliptically deformed confinement potential as in Ref. \cite{fuhrer} in particular.

The gating techniques for phosphorene have been developped \cite{l20,l21,l22} and applied for fabrication
of the field-effect transistors. 
The ring-like potential can be defined electrostatically in a plane plate capacitor system
with a tubular electrode protruding from one of the plate electrodes with the phosphorene 
layer embedded in a dielectric \cite{apl,apl2}. For a similar gating system defined for electrons on liquid helium surface, see e.g. Ref. \cite{szafran2}.

The results of this paper indicate the way to observe the Aharonov-Bohm effect for the system confined by the external
potential in phosphorene. For circular quantum rings the Aharonov-Bohm oscillations 
can only be observed  in the excited part of the spectrum since in the degenerate ground state the electron density forms separated islands
and the persistent current circulation is interrupted. For an elliptical deformation of the confinement potential,
the oscillations of the ground-state parity appear with the Aharonov-Bohm periodicity.
The amplitude of these oscillations has been determined (see $\Delta_0$ in Table II).
Based on results of Ref. \cite{fuhrer} one can expect that for $\alpha\leq 1.2 \frac{m_x}{m_y}$ the ground-state oscillation should enter the experimental resolution.

 For a specifically chosen eccentricity parameter $\alpha$ one can 
reduce the spectra to those that are characteristic to a circular quantum ring with an isotropic effective mass. 
We explained this effect analytically in the effective mass approximation that indicates an additional symmetry found for 
a value of $\alpha$ and the spectra that agree with the tight-binding ones
up to an avoided crossing in the excited energy spectra found in the latter,
which is minimal for the optimal value of $\alpha=\frac{m_x}{m_y}$ (see $\Delta_e$ in Table II).
The width of the avoided crossing should also be accessible for an experimental study.

\section{Summary and Conclusions}
We have studied the ground-state energy oscillations in a quantum ring potential
defined within monolayer black phosphorus with the tight-binding and effective mass
models. In a circular quantum ring, the strong anisotropy of the effective mass produces a ground state localized along the axis related to the heavier mass. The current
circulation around is possible for an elliptic ring.
A braided pattern of even and odd-parity energy levels is then observed in the ground state
with crossings appearing with the Aharonov-Bohm periodicity.
In particular, for the ellipse with the ratio of the semiaxes equal to the effective masses ratio, the electron density becomes uniform along the ring. Then, 
the single-electron energy spectrum becomes similar to that of a circular quantum ring
with an isotropic effective mass equal to the geometric average of the effective masses
along the two crystal directions. We demonstrated that the angular momentum in the rescaled space $l_z'$ is definite in the single-electron Hamiltonian eigenstates. We provided an analytical formula for the spectrum in the 1D limit.
The appicability of the $l'_z$ operator exceeds the quantum rings and can be used for modeling  other confined systems in phoshorene.
For two electrons, an elliptical deformation of the ring  produces avoided crossings
in the ground-state due to the same parity of low-energy spin-polarized states
appearing periodically on the magnetic field scale.

\section*{Acknowledgments}
This work was supported by the National Science Centre
(NCN) according to decision DEC-2019/35/O/ST3/00097.
Calculations were performed on the PLGrid infrastructure.


\begin{thebibliography}{00}
\bibitem{reference} With $\alpha=\frac{m_x}{m_y}$ for $R_s$=25, 30, 35, \dots, 70 (in nm)
we obtain the lowest conduction band energy level at
341.066,  340.764, 340.559, 340.415,  340.328,  340.266, 340.184,
340.161, and 340.138 (in meV), respectively.
\end{thebibliography}
\end{document}